# Using skateboarding to develop a culturally relevant tutorial on static equilibrium

Gian Viray[*], Isaac Cheney[†], and Tong Wan[†a]

[*]Burnett School of Biomedical Sciences, University of Central Florida, 4364 Scorpius Street, Orlando, FL 32816, USA

[†]Department of Physics, University of Central Florida, 4111 Libra Drive, Orlando, FL 32816, USA

Culturally relevant pedagogy (CRP),[1] initially developed by Ladson-Billings, is an instructional framework for supporting diverse learners by drawing on their cultural backgrounds and experiences. In line with the CRP framework, we developed a tutorial on static equilibrium using skateboarding—a popular activity on university campuses—as a culturally relevant context. To help students refine their conceptions about static equilibrium documented in the physics education research (PER) literature, we used the elicit-confront-resolve (ECR) strategy to develop the tutorial.[2,3] In this paper, we provide a detailed account of how we operationalized the ECR strategy in designing the sequences of questions in the tutorial. Additionally, we present anecdotal evidence to show that this research-based culturally relevant tutorial appears to effectively engage students and motivate their interest in learning physics.

## 1 Culturally Relevant Pedagogy

The CRP framework emphasizes students' cultural references as starting points for teachers to support student learning. It is suggested that student learning needs to be contextualized by relating content to students' real-life experiences. Recent studies have demonstrated specific ways to enact CRP in teaching physics at the K-12 level, which increased students' interest and improved their attitude toward physics.[4,5] For example, a high school teacher tried to connect students' interest in Disney to learning energy efficiency by having students work on a project of finding the best route to drive to Disney World. CRP has also been suggested as a useful model in college-level physics instruction for supporting diverse learners.[6] However, research in implementing CRP in undergraduate physics courses remains scarce.

To motivate students' interest and increase engagement in learning physics, we use a skateboarding scenario to develop a culturally relevant tutorial on static equilibrium. Skateboarding is a popular action sport and a recreational activity with a large community. According to a 2017 national survey, there are approximately 6.3 million participants in the US, with 4.7 million aged 24 years or younger.[7] Moreover, skateboarding is well received among college students as a sport or a means of transportation, especially in the southeastern US. At a metropolitan public research university, skateboards are commonly used for students to get around campus. Additionally, the first two authors have been active members of the skateboarding club at the university since their first year here.[8] Given the popularity of skateboarding among college students, we believe skateboarding is a great topic to use for creating cultural connections to physics concepts in introductory physics courses.

## 2 Student Conceptions of Static Equilibrium

Prior research has shown that static equilibrium is a challenging topic for both high school and university students.[9-11] In introductory-level undergraduate physics courses, Ortiz et al. identified multiple student conceptions about the relationship of force and torque to a system in static equilibrium as well as conceptions about the center of mass.[10]

The tutorial we developed is intended to help students refine the following conceptions documented in Ortiz et al. When students considered a balanced system with a continuous mass distribution, students often assumed that the forces applied on both sides of a fulcrum must be equal. When given a tilted but balanced system, students' responses were consistent with a belief that the tilted orientation is caused by unbalanced torques or forces. Additionally, many students gave responses that were consistent with a belief that the center of mass divides a system into two components with equal mass.

[a] Tong.Wan@ucf.edu

## 3   The Elicit-Confront-Resolve Strategy

ECR is a common instructional strategy used in guided inquiry to support students in refining their conceptions (also referred to as "difficulties") to more strongly align with normative physics. In the *elicit* phase, known conceptions that learners have are drawn out using a physics task, typically the same as the one used in the original research for identifying student conceptions. In the *confront* phase, students are guided through a reasoning process that helps them become consciously aware of specific flaws in their thinking. Often, the question sequence leads students to recognize an inconsistency or a contradiction between two competing ideas or lines of reasoning. Lastly, in the *resolve* phase, scaffolding is provided to support students to develop ideas and reasoning that are aligned with normative physics.

## 4   Design of Tutorial

The tutorial consists of three sections. The first section includes a modified version of two questions about a two-piece bar from Ortiz et al.[10] The two questions can be considered the *central* questions of this tutorial used to elicit student ideas about static equilibrium. The second and third sections center around a skateboarding trick, called the "manual". The skateboarding scenario is used to help students relate the concepts about static equilibrium to their everyday experiences and reconcile their intuitive ideas with the concepts. Both sections in the skateboarding context start with an elicit question analogous to the relevant two-piece bar question. It is then followed by a sequence of questions intended to help students confront and resolve inconsistencies in their thinking about the skateboarding elicit questions. At the end of both sections, students are guided to revisit the two-piece bar questions and resolve inconsistencies. A summary of the questions used in each of the ECR phases, and their aims is shown in Table I. Below, we describe the questions in detail.

Table I. A summary of the tutorial questions used in each of the ECR phases and their aims.

| Section | ECR phase | Question(s) | Aim of question(s) |
|---|---|---|---|
| I | Elicit | Weight comparison question in the two-piece bar context | To draw out incorrect ideas, such as "a tilted orientation is caused by an unbalanced torques or forces" |
| | Elicit | Center of mass location question in the two-piece bar context | To draw out incorrect ideas, such as "the center of mass divides an object into two pieces of equal mass" |
| II | Elicit | Force comparison question in the skateboarding context | To draw out student ideas about balancing on a tilted skateboard |
| | Confront | A sequence of questions in the skateboarding context related to net force, free body diagram, lever arm, and net torque | To guide students to reason about the lever arm and net torque in the skateboarding context |
| | Confront and Resolve | Fictious student dialogue question in the skateboarding context | To help students recognize the inconsistency between "the forces are equal" and "the net torque is zero" by accounting for the lever arms |
| | Resolve | Revisit of the two-piece bar weight comparison question | To help students resolve inconsistency by drawing free body diagram that includes forces and lever arms |

| III | Elicit | Center of mass location of the skater-skateboard system question | To draw out student ideas about the location of the center of mass for the skater-skateboard system |
| | Confront | A sequence of questions in the skateboarding context related to free body diagram for the skater-skateboard system, torque, and net torque | To help students recognize the contradiction between "the center of mass is over the back foot" and "the net torque is zero" |
| | Resolve | Revisit of the skater-skateboard center of mass location question | To help students justify why the center of mass of the system must be directly over the pivot |
| | Resolve | Revisit of the two-piece bar center of mass location question | To help students relate static equilibrium to the center of mass location of the bar |

In section I, students consider a two-piece bar problem as shown in Fig. 1. The bar has shaded and unshaded pieces, which are made with different materials. The bar is at rest on a frictionless pivot. On the first question, students are asked if the weight of the shaded piece is greater than, less than, or equal to the weight of the unshaded piece. On the second question, students are asked to identify the marked point (A, B, or C) that would best represent the location of the center of mass of the bar. Students are asked to explain their reasoning on both questions. We expect that these questions would elicit some of the incorrect ideas documented in Ortiz et al, such as "a tilted orientation is caused by an unbalanced torques or forces", and "the center of mass divides an object into two pieces of equal mass".[10]

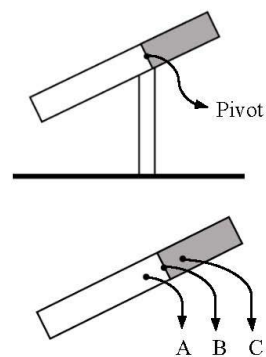

Fig. 1. Section I of the tutorial with a modified version of the two-piece bar problem from Ortiz et al.[10]

In section II, students consider the "manual", a skateboarding trick, as shown in Fig. 2. A skater stands on a skateboard with only the back wheels in contact with the ground. The first question asks students if the force on the skateboard by the front foot (i.e., the one close to front wheels) is greater than, less than, or equal to the force by the back foot. This question is analogous to the weight comparison question of the two-piece bar problem, and it is used to elicit students' intuitive ideas about the manual trick. We expect that some students are able to draw on their experience and conclude that the back foot exerts a greater force; others may believe that equal forces by both feet are required for static equilibrium, consistent with a student belief identified by Ortiz et al.

II. A skater is performing a trick called a manual as shown in the figure. The front wheels and the tail of the board are in the air, and only the back wheels are touching the ground. Assume that the skater manages to stay at rest for a few seconds. The skateboard has a uniform density, and the wheels are frictionless.
   A. During the manual, is the force exerted on the skateboard by the skater's front foot (i.e., the one close to the front wheels) greater than, less than, or equal to the force exerted by the back foot? Explain your reasoning.

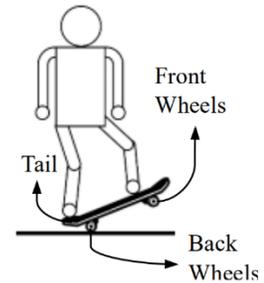

Fig. 2. Excerpt from Section II of the tutorial.

The elicit question is then followed by a sequence of questions to guide students to confront the incorrect idea that the forces by front foot and back foot are equal. First, students are asked for the net force on the skateboard. Students are then asked to draw an extended free body diagram for the skateboard, and to sketch the lever arm for each force. Then, students are asked for the net torque on the skateboard. We expect that most of the students would reason that the net torque is zero because the skateboard is in static equilibrium.

To further engage students in a deliberate examination of competing lines of reasoning and resolve inconsistencies in their own thinking, we task students to consider a fictitious dialogue of three students. Student 1 argues that since the net torque is zero, the forces by front and back feet must be equal. Student 2 discusses the relationships among the torques by front foot, back foot, and gravity of the skateboard, compares the lever arms, and then correctly concludes that force by the back foot is greater. Student 3 correctly states that force by the back foot is greater but incorrectly states that this is because the back foot is below the pivot and then concludes that lever arm does not matter. We hope students would realize that the forces by the feet cannot be equal because the force by the front foot has a greater lever arm than the force by the back foot does, inconsistent with an established condition that the net torque is zero. Through identifying and articulating the flaws in the statements of students 1 and 3, we hope students would be able to resolve inconsistency and realize that both the lever arm and force should be accounted for when considering the torque.

At the end of section II, students are prompted to revisit the two-piece bar problem and whether they would change their answer. To scaffold resolving inconsistency, students are asked to draw a free body diagram including forces and lever arms and explain how the lever arms they sketched help them compare the forces.

The third section of the tutorial concerns the center of mass (c.m.) location in the skateboarding context. Students are told to consider the skateboard and the skater together as a system. The first question serves as an elicit question on which students are asked for the location of the c.m. of the system. This question is analogous to the second question in the two-piece bar scenario. We expect that many students would argue that the c.m. is to the left of the pivot because the force by the back foot is greater.

In the confront phase, students are first asked to draw a free body diagram for the skater-skateboard system. They then consider an example free body diagram, which incorrectly labels the c.m. directly over the back foot (see Fig. 3). Students are prompted to consider the torque by each force and the net torque before they are asked whether the diagram is correct. The goal is to help students recognize the contradiction between the c.m. being directly over the back foot and the net torque of the system being zero.

C. Now consider a free body diagram of the system sketched by a classmate as shown. In the diagram, $CM_{sys}$ represents the center of mass of the system, $\vec{F}_g$ represents the gravitational force of the system, and $\vec{N}$ represents the normal force by the ground.

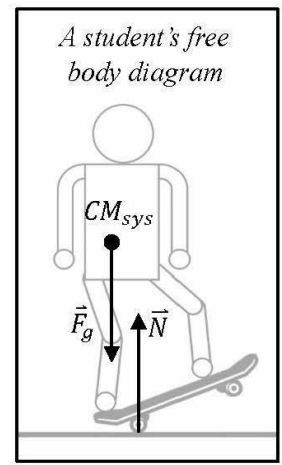

*A student's free body diagram*

  i. According to this diagram, is the torque produced by the gravitational force zero or non-zero? Explain your reasoning.

  ii. According to this diagram, is the torque produced by the normal force zero or non-zero? Explain your reasoning.

  iii. According to this diagram, is the net torque zero or non-zero? Explain your reasoning.

  iv. Based on your answers to the questions above, do you think this diagram is correct or incorrect? Explain your reasoning.

Fig. 3. Excerpt from Section III of the tutorial.

Students are then asked where the c.m. must be located given that the net torque on the system is zero. This question guides students to resolve the inconsistency identified previously. We hope students would realize that the c.m. must be directly over the pivot so that the net torque on the system is zero.

Lastly, students are prompted to revisit the question about the location of c.m. of the two-piece bar and resolve inconsistency. They are asked whether the bar would rotate or remain at rest if the c.m. of the bar is not directly over the pivot.

## 5   Student reception of the tutorial

The tutorial was administered in a calculus-based introductory physics course that focuses on mechanics at the university. The course was taught in studio mode, which integrates lecture, tutorial, and labs. Most of the students were science and engineering majors. Students worked through the tutorial in groups, facilitated by an instructor, a graduate teaching assistant, and an undergraduate learning assistant.

During the tutorial, students appeared motivated and engaged. Many students drew on their daily experiences and intuition during discussion. Interestingly, one student—a likely frequent skater— brought their own skateboard into the classroom that day and used it to demonstrate the trick. Moreover, the student encouraged others without skateboarding experience to try and "get a feel" for the forces and the location of c.m. Overall, the anecdotal evidence seems to suggest that the tutorial provides students with a valuable opportunity to apply physics concepts in a relatable real-world context.

## 6   Conclusion

In this article, we have provided an illustrative example of how curriculum designers can operationalize the ECR strategy to develop a culturally relevant tutorial. Our informal classroom observation suggested that this research-based tutorial appeared to engage students and motivate their interest in learning physics. Future work is needed to validate the tutorial by surveying students' motivations and examining their performance on questions about static equilibrium.